\documentstyle[12pt]{article}

\newlength{\extraspace}
\newlength{\extraspaces}
\newcommand{\be}{\begin{equation}
\addtolength{\abovedisplayskip}{\extraspaces}
\addtolength{\belowdisplayskip}{\extraspaces}
\addtolength{\abovedisplayshortskip}{\extraspace}
\addtolength{\belowdisplayshortskip}{\extraspace}}
\newcommand{\ee}{\end{equation}}
\newcommand{\ba}{\begin{eqnarray}
\addtolength{\abovedisplayskip}{\extraspaces}
\addtolength{\belowdisplayskip}{\extraspaces}
\addtolength{\abovedisplayshortskip}{\extraspace}
\addtolength{\belowdisplayshortskip}{\extraspace}}
\newcommand{\ea}{\end{eqnarray}}
\newcommand{\nonu}{\nonumber \\[.5mm]}

\newcommand{\e}{\, {\rm e}}
\newcommand{\D}{{\cal D}}
\newcommand{\bra}[1]{\left\langle {#1} \right\vert}
\newcommand{\ket}[1]{\left\vert {#1} \right\rangle}
\newcommand{\VEV}[1]{\left\langle {#1} \right\rangle}

\newcommand{\ob}[1]{{\cal{O}}_{#1}}
\newcommand{\sst}{\scriptstyle}

\newcommand{\ra}{\rightarrow}

\newcommand{\lra}{\leftrightarrow}

\newcommand{\del}{\partial}
\newcommand{\half}{{\textstyle{1\over 2}}}

\newcommand{\ghat}{\hat{g}}
\newcommand{\sqghat}{\sqrt{\ghat}}

\newcommand{\PSI}[1]{\psi_{-(#1+\half)}}
\newcommand{\BPSI}[2]{\bar{\psi}_{#1 #2+\half}}
\newcommand{\AL}[1]{\alpha_{#1}}
\newcommand{\bino}[2]{
\left(\!
\begin{array}{c}
	#1 \\ #2
\end{array}
\!\right)
}
\begin{document}
\addtolength{\baselineskip}{.5mm}
\thispagestyle{empty}
%
%%%%%%%%%%%%%%%%%%%%%%%%%%%%%%%%%%%%%%%%%%%%%%%%%%%%%%%%%%%
%                      Title  Page                        %
%%%%%%%%%%%%%%%%%%%%%%%%%%%%%%%%%%%%%%%%%%%%%%%%%%%%%%%%%%%
\begin{flushright}
TIT/HEP--220 \\
April, 1993
\end{flushright}
\vspace{3mm}
\begin{center}
{\Large{\bf{ Factorization in $2D$ String Theory}}} \\[24mm]
{\sc Hiroshi Shirokura}\footnote{e-mail: siro@phys.titech.ac.jp} \\[8mm]
{\it Department of Physics, Tokyo Institute of Technology \\[3mm]
Oh-okayama, Meguro, Tokyo 152, Japan} \\[29mm]
%
%%%%%%%%%%%%%%%%%%%%%%%%%%%%%%%%%%%%%%%
%            Abstract                 %
%%%%%%%%%%%%%%%%%%%%%%%%%%%%%%%%%%%%%%%
{\bf Abstract}\\[9mm]
{\parbox{13cm}{\hspace{5mm}
We show the factorization of correlation functions of tachyon operators
in $2D$ string theory using the discretized approach.
Our results can be understood in terms of the operator product expansion
of tachyon operators.
We also give a systematic way of computing correlation functions of
tachyon operators.
}}
\end{center}
\vfill
\newpage
\setcounter{equation}{0}
%
%%%%%%%%%%%%%%%%% Introduction %%%%%%%%%%%%%%%%%%%%%%%%%%%%%%%%%%%%%%%%%%%%

Developments in the last few years in two-dimensional quantum gravity
coupled to $c\leq 1$ conformal field theories \cite{BIPZ}-\cite{POLREV}
have cast new light on theories of gravity in higher dimensions.
The most interesting is the $c=1$ Liouville theory which can be regarded
as a critical string theory in a two-dimensional target space.
Tachyon correlation functions have been studied by many people
from the point of view of continuum approach
\cite{GOULI}-\cite{SATAPART}.
Sakai and Tanii demonstrated the factorization of the $N$-point functions in
the $c=1$ Liouville theory \cite{SATAFACT}.
They understood the factorization as a result of the short-distance
singularity
arising from the operator product expansion of two tachyon operators.
% Their results is valid only when the number of cosmological term insertion
% $s$ is a non-negative integer.
% These restriction come mainly from the existence of the limit of calculation
% technique in the path integral approach.
Because of the limitations of the path-integral technique, their argument was
restricted to a particular kinematical situation.

On the other hand, Dijkgraaf, Moore, and Plesser showed that
the tachyon amplitudes can be reformulated in terms of a scattering
process for free fermions \cite{DMP}.
They derived a compact expression for the generating functional of
the correlation functions of tachyon operators. Their generating functional
can naturally be considered a tau function of the KP hierarchy.
It is very natural to apply their powerful method to the problem of
factorization.

The purpose of the present article is to provide a more general demonstration
of the factorization than in the paper of Sakai and Tanii.
We find that the factorization is independent of the topology of the
two-dimensional manifold.
We can understand this from the local nature of the operator product
expansion of tachyon operators.
We calculate correlation functions of tachyon operators
explicitly for some simple cases and confirm that these correlation
functions indeed satisfy our factorization rule.

%%%%%%%%%%%%%%%  Path integral approach %%%%%%%%%%%%%%%%%%%%%%%%%%

In this article we follow the notation of refs.\cite{SATAPART},
\cite{SATAFACT}.
The $c=1$ Liouville theory is described by the action
\ba
S[\phi, X; \mu]   & = & S_{\rm Liouville} + S_{\rm matter} \,, \nonu
S_{\rm Liouville} & = & \frac{1}{8\pi}\int\!\!d^2z\sqghat
  \biggl(\ghat^{ab}\del_a\phi\del_b\phi-2\sqrt{2}\hat{R}\phi
  +8\mu\,\e^{-\sqrt{2}\phi}\biggr)\,, \nonu
S_{\rm matter}    & = & \frac{1}{8\pi}\int\!\!d^2z\sqghat
  \ghat^{ab}\del_a X\del_b X\,,
\ea
where $\phi$ is the Liouville field and $X$ a massless scalar field
with central charge $c=1$.
The physical states in the $c=1$ Liouville theory are well-known from
the study of BRST cohomology \cite{LIAN}. We have two kinds of states:
tachyon states and boundary states.
Now we concentrate on the tachyon states.
We shall use the Euclidean signature spacetime.
In the $c=1$ Liouville theory, the tachyon operators are given
in the form \cite{SATAFACT}
\be
\ob{p}
= \int\!\!d^2z\sqghat\,{\cal{T}}_p(z)
= \int\!\!d^2z\sqghat\,
\e^{ipX}\,
\e^{\beta(p)\phi}
\,,
\label{eqn:tachyon}
\ee
where $p$ is the momentum parameter of the tachyon.
 From BRST invariance we have the on-shell condition
\be
\half p^2-\half\beta(\beta+2\sqrt{2})= 1\,.
\label{eqn:on-shell}
\ee
In spite of the name {\it tachyon}, this state corresponds to a massless ground
state in the language of string theory in a two-dimensional target space
\cite{DFKU}.
We can easily understand this statement as follows.
The relation between the tachyon operator
${\cal{T}}_p$ and the wave function of the corresponding state is given by
${\cal{T}}_p=g_{\rm st}\Psi(\phi, X)$,
where $g_{\rm st}$ is the string coupling constant. If we recall that
$g_{\rm st}\propto\exp(-\sqrt{2}\phi)$, the wave function has the following
form
\be
\Psi(\phi, X) = \exp
\left[
	ipX+
	\left(
		\beta+\sqrt{2}
	\right)\phi
\right]\,.
\ee
We thus interpret the Liouville field as Euclidean time and regard
the energy of this state as $E=\beta+\sqrt{2}$. We can rewrite
(\ref{eqn:on-shell}) in the following form
\be
E^2=p^2+m^2\,,\quad m^2=0\,.
\ee

Now let us briefly discuss the factorization of the tachyon amplitudes
in the continuum approach.
The $N$-point correlation function of tachyon operators is
given by a path integral \cite{SATAPART}
\ba
\VEV{\ob{p_1}\cdots\ob{p_N}} &\!\!\! = \!\!\!&
\int\!\!\frac{\D\phi\D X}{V_{\rm CKV}}\,\e^{-S[\phi, X; \mu]}\,
\ob{p_1}\cdots\ob{p_N} \nonu
&\!\!\! = \!\!\!& 2\pi\delta
\left(
	\sum^N_{k=1}p_k
\right)
\,\frac{\Gamma(-s)}{\sqrt{2}}
\tilde{A}(p_1,\ldots,p_n)\,,
\label{eqn:N-amplitude}
\ea
\ba
\tilde{A} & = & \int\!\!\frac{[d\tau]}{V_{\rm CKV}}
\prod^N_{k=1}\,\int\!\!d^2z_k\sqrt{\ghat(z_k)}\,
\VEV{\prod^N_{k=1}{\e^{ip_kX(z_k)}}}_{\tilde{X}} \nonu
&  & \times
\VEV{
\prod^N_{k=1}\,\e^{\beta(p_k)\phi(z_k)}
\left(
	\frac{\mu}{\pi}\int\!\!d^2w\sqrt{\ghat(w)}\,
	\e^{-\sqrt{2}\phi(w)}
\right)^s
}_{\tilde{\phi}}\,,
\ea
\[ s = 2-2h-N+\frac{1}{\sqrt{2}}\sum^N_{k=1}\,|p_k|\,, \]
where $s$ is the number of the cosmological term insertion
and the volume $V_{\rm CKV}$ is that generated by conformal Killing vectors.

In the $c=1$ Liouville theory, the notion of {\it chirality} appears
\cite{SATAPART}.
As a solution of (\ref{eqn:on-shell}) we choose
$\beta(p)=-\sqrt{2}+|p|$ following the argument of
refs.\cite{SEIREV},\cite{POLREV}.
We define the chirality of the tachyon operator (\ref{eqn:tachyon})
to be positive if $p > 0$ and to be negative if $p < 0$.
Any zero-momentum tachyon operators decouple in (\ref{eqn:N-amplitude})
and make the path integral vanish.
In the sphere topology, the integral for non-zero modes can be explicitly
performed after fixing the $SL(2, \mbox{\boldmath$C$})$ gauge.
In particular, if $p_1 < 0, p_2,\ldots,p_N > 0$,
the amplitude is given in a form that allows analytic continuation
with respect to $s$
\ba
\VEV{\ob{p_1}^{(-)}\ob{p_2}^{(+)}\cdots\ob{p_N}^{(+)}}_{h=0}
 & \equiv & A(p_1^{(-)},p_2^{(+)},\ldots,p_N^{(+)}) \nonu
 & = & \pi^{N-3}\mu^s\frac{\Gamma(-s)}{\Gamma(N+s-2)}\,
\prod^N_{k=2}\,\Delta(1-\sqrt{2}|p_k|),
\label{eqn:sata-amplitude}
\ea
where $s=2-N+\sqrt{2}|p_1|$ and $\Delta(x)\equiv\Gamma(x)/\Gamma(1-x)$.
In (\ref{eqn:sata-amplitude}), we redefined the $N$-point function
as follows
\be
\VEV{\ob{p_1}\cdots\ob{p_N}}_{h=0}\equiv
\Gamma(-s)\,\tilde{A}(p_1,\ldots,p_N)\,.
\ee

As was discussed in \cite{SATAFACT}, all the singularities in
$p_k\,,(k=2,\ldots,N)$ come from the short-distance singularities between
${\cal{T}}_{p_1}(z)$ and ${\cal{T}}_{p_k}(0)$,
\ba
\lefteqn{:\e^{ip_1X}\e^{\beta(p_1)\phi}(z):
         :\e^{ip_kX}\e^{\beta(p_k)\phi}(0):}
\nonu
& \sim & \sum^{\infty}_{n=0}\left(\frac{1}{n!}\right)^2
|z|^{2\vec{p_1}\cdot\vec{p_k}+2n}
:\e^{ip_1X}\e^{\beta(p_1)\phi}(0)
\del^n\bar{\del}^n\e^{ip_kX}\e^{\beta(p_k)\phi}(0):\,,
\label{eqn:OPE}
\ea
where $\vec{p}_i = (p_i ,-i\beta(p_i))$.
The operator that appears in the $n=0$ term of the right-hand side of
(\ref{eqn:OPE}) has the same form as a tachyon operator and its
corresponding state has the following mass
\ba
M^2 & = & (\beta(p_1)+\beta(p_k)+\sqrt{2})^2-(p_1+p_k)^2 \nonu
    & = & -(2\vec{p_1}\cdot\vec{p_k}+2) \nonu
    & = & 2(1+\sqrt{2}p_1)(1-\sqrt{2}p_k)\,.
\ea
Therefore the pole at $p_k = 1/\sqrt{2}\ (n=0)$ is due to a tachyon
intermediate state.
In the case of (\ref{eqn:sata-amplitude}), the residue of the pole at
$p_k=1/\sqrt{2}$
is indeed given by a $(N-1)$-point tachyon amplitude
\be
A(p_1,p_2,\ldots,p_N)
\stackrel{p_k\ra 1/\sqrt{2}}{\approx}
-\frac{\pi}{(1+\sqrt{2}p_1)(1-\sqrt{2}p_k)}\,
A(p,p_2,\ldots,p_{k-1},p_{k+1},\ldots,p_N)\,,
\ee
where $p = p_1+1/\sqrt{2}$ is the momentum of the intermediate
tachyon operator whose chirality must be negative.

 From the above mentioned argument, all we have to do to study the
factorization
of the tachyon amplitudes (\ref{eqn:N-amplitude}) is take the limit of
$p_k \ra 1/\sqrt{2}\ (p_k\ra -1/\sqrt{2})$ for a tachyon operator
${\cal{T}}_{p_k}$ with positive (negative) chirality and confirm that
the residue of the pole is indeed that given by the $(N-1)$-point tachyon
amplitude.
To do this the discrete approach of Dijkgraaf, Moore, and Plesser is more
convenient than the path integral approach of Sakai and Tanii.

%%%%%%%%%%%%%%% DMP's approach to tachyon amplitudes %%%%%%%%%%%%%%%%

Dijkgraaf, Moore, and Plesser
expressed the tachyon amplitudes (\ref{eqn:N-amplitude}) as
vacuum expectation values in a conformal field theory \cite{DMP}.
We adjust their result to fit the results of the continuum theory
($X$ field is compactified in radius $\beta$)
\ba
\lefteqn{A(\ob{p_1}^{(-)},\ldots,\ob{p_m}^{(-)},
\ob{p_{m+1}}^{(+)},\ldots,\ob{p_N}^{(+)})} \nonu
& \equiv & \lim_{\beta\ra\infty}(-i)^N\pi^{N-3}\frac{1}{\beta}
 \prod^N_{j=1}\frac{\Gamma(-n_j/\beta)}{\Gamma(n_j/\beta)}
 \mu^{\frac{1}{2\beta}\sum^N_{j=1}n_j} \nonu
&    & \times \bra{0}\AL{n_N}\cdots\AL{n_{m+1}}
 S\AL{-n_{m}}\cdots\AL{-n_1}\ket{0}\,,
\label{eqn:dmp-formula}
\ea
where $\alpha$ denotes the creation or annihilation operator of a free boson
and $\ket{0}$ is the standard $SL(2, \mbox{\boldmath$C$})$ invariant vacuum.
The correspondence between the momentum parameters of the tachyons
and the subscripts of the $\alpha$ modes is given by
\ba
\sqrt{2}p_k &\!\! =      -n_k/\beta\,, & \qquad(\mbox{for}\quad k=1,\ldots,m)
\,,\nonu
\sqrt{2}p_l &\!\! = \ \ \ n_l/\beta\,, & \qquad(\mbox{for}\quad l=m+1,\ldots,N)
\,.
\ea
The scattering matrix $S$ of the scattering process discussed in \cite{MOORE}
is obtained using matrix quantum mechanics.
It is described in terms of fermion modes
\be
\AL{n}=\sum_m\PSI{m}\BPSI{n}{+m}\,,\quad
\Bigl\{\PSI{m},\BPSI{}{n}\Bigr\} = \delta_{n+m,0}\,,
\ee
as follows
\be
S=:\mbox{exp}\,
\left[
	\sum_m\Bigl(\log R_{p_m}\Bigr)
	\PSI{m}\BPSI{}{m}
\right]
:\,,
\label{eqn:S-matrix}
\ee
where $\sqrt{2}p_m=\frac{1}{\beta}\Bigl(m+\half\Bigr)$ is the momentum of
the loop fermion \cite{MOORE} which is discretized as a result of
the compactification.
In (\ref{eqn:S-matrix}), $R_{p}$ is the reflection coefficient of
the scattering process
\be
R_{p} = \e^{i\mu\log\mu}\mu^{-\sqrt{2}p}
\sqrt{\frac{\Gamma(\half-i\mu+\sqrt{2}p)}{\Gamma(\half+i\mu-\sqrt{2}p)}}\,.
\label{eqn:reflection}
\ee
Note that (\ref{eqn:reflection}) is valid only for sufficiently large $\mu$.
As the formula (\ref{eqn:dmp-formula}) has its origin in matrix quantum
mechanics, it has the form of a sum over all topologies.
One can expand it in terms of the genus as follows \cite{DDK}, \cite{DMP}
\ba
\lefteqn{A(\ob{p_1}^{(-)},\ldots,\ob{p_m}^{(-)},
\ob{p_{m+1}}^{(+)},\ldots,\ob{p_N}^{(+)})} \nonu
& = & \sum^N_{h=0} \mu^{-2h}
A(\ob{p_1}^{(-)},\ldots,\ob{p_m}^{(-)},
\ob{p_{m+1}}^{(+)},\ldots,\ob{p_N}^{(+)})_h\,.
\label{eqn:genus-expansion}
\ea
This asymptotic expansion is only true for large cosmological constant
$\mu$.

%%%%%%%%%%%% Explicit calculation of tachyon amplitudes %%%%%%%%%%%%%%%%

Now let us expand (\ref{eqn:dmp-formula}) in terms of the genus and perform
an explicit computation of the tachyon amplitudes for some cases.
 First we rewrite it as follows
\ba
\lefteqn{A(p_1^{(-)},\ldots,p_m^{(-)},
p_{m+1}^{(+)},\ldots,p_N^{(+)})} \nonu
& = \!\!& \lim_{\beta\ra\infty}(-i)^N\pi^{N-3}\frac{1}{\beta}
 \prod^N_{j=1}\frac{\Gamma(-n_j/\beta)}{\Gamma(n_j/\beta)}
 \mu^{\frac{1}{2\beta}\sum^N_{j=1}n_j} \nonu
&    \!\!& \times \bra{0}\AL{n_N}\cdots\AL{n_{m+1}}
 \prod^{m-1}_{k=0}(S\AL{-n_{m-k}}S^{-1})\ket{0} \nonu
& = \!\!& \lim_{\beta\ra\infty}(-i)^N\pi^{N-3}\frac{1}{\beta}
 \prod^N_{j=1}\frac{\Gamma(-n_j/\beta)}{\Gamma(n_j/\beta)}
 \mu^{\frac{1}{2\beta}\sum^N_{j=1}n_j} \nonu
&    \!\!& \times \bra{0}\AL{n_N}\cdots\AL{n_{m+1}}
 \prod^{m-1}_{k=0}\sum_nR_{p_n}
 R^*_{n_{m-k}/\sqrt{2}\beta-p_n}\PSI{n}\BPSI{-n_{m-k}}{+n}\ket{0}.
\label{eqn:fermionization}
\ea
We can expand $RR^*$ in (\ref{eqn:fermionization}) using the following
asymptotic expansion derived from the Stirling formula,
\be
z^{b-a}\frac{\Gamma(z+a)}{\Gamma(z+b)}
\approx \sum^\infty_{n=0}\bino{B}{n}
\sum^{[n/2]}_{m=0}(-1)^mI_m(B)\frac{d^{2m}}{dA^{2m}}
\left(
	\frac{A}{z}
\right)^n\ \ \mbox{as}\ z\ra\infty\,,
\ee
\[
A=(a+b-1)/2,\quad B=a-b\,,
\]
where the $I_m(x)$ is a polynomial of $x$ of $m$-th order and the first few
polynomials are given by
\ba
&  & I_0 = 1\,,\quad, I_1=\frac{1}{4!}(2x+1)\,,\quad
I_2=\frac{1}{8\cdot 6!}(2x+1)(10x+7)\,, \nonu
&  & \cdots \ .
\ea
Applying it twice in $RR^*$, we get the following asymptotic expansion,
\be
R_{p_m}R^*_{n/\sqrt{2}\beta-p_m}\approx
\sum^\infty_{k=0}\bino{n/\beta}{k}
\sum^{[k/2]}_{l=0}(-1)^l(2l)!\bino{k}{2l}I_l(n/\beta)
\left(
	m+\frac{1}{2}-\frac{n}{2}
\right)^{k-2l}\!\!
\left(
	\frac{i}{\beta\mu}
\right)^k.
\label{eqn:RR-star}
\ee
 From eqs.(\ref{eqn:fermionization}),(\ref{eqn:RR-star})
we can expand the amplitude (\ref{eqn:dmp-formula}) in terms of $\mu^{-1}$.
We can compute each term of (\ref{eqn:genus-expansion}), since it is
only a vacuum expectation value in a conformal field theory.
We succeed in summarizing the results of the computation in compact form for
the following three cases.

\begin{description}
\item[\rm Case 1.]$h=0,\ (-,+,\ldots,+)\,,\quad (s = \sqrt{2}|p_1|-N+2)$,
\ba
\lefteqn{A(p_1^{(-)},p_2^{(+)},\ldots,p_N^{(+)})_{h=0}} \nonu
& = & \lim_{\beta\ra\infty}(-i)^N\pi^{N-3}\frac{1}{\beta}
 \prod^N_{j=1}\frac{\Gamma(-n_j/\beta)}{\Gamma(n_j/\beta)}
 \mu^{\frac{1}{2\beta}\sum^N_{j=1}n_j} \nonu
&   & \times \bra{0}\AL{n_N}\cdots\AL{n_2}
 \sum_m\left(\frac{i}{\beta\mu}\right)^{N-2}
 \bino{n_1/\beta}{N-2}m^{N-2}\PSI{m}\BPSI{-n_1}{+m}\ket{0} \nonu
& = & \lim_{\beta\ra\infty} -\pi^{N-3}
 \prod^N_{j=1}\frac{\Gamma(-n_j/\beta)}{\Gamma(n_j/\beta)}
 \mu^{n_1/\beta+2-N}
 \frac{\Gamma(n_1/\beta+1)}{\Gamma(n_1/\beta-N+3)}
 \frac{n_2}{\beta}\frac{n_3}{\beta}\cdots\frac{n_N}{\beta} \nonu
& = & \mu^s\frac{\Gamma(-s)}{\Gamma(N+s-2)}\prod^N_{k=2}
 \Delta(1-\sqrt{2}p_k)\,.
\ea
\item[\rm Case 2.]$h=1,\ (-,+,\ldots,+)\,,\quad (s = \sqrt{2}|p_1|-N)$,
\ba
\lefteqn{A(p_1^{(-)},p_2^{(+)},\ldots,p_N^{(+)})_{h=1}} \nonu
& = & -\frac{\pi^{N-3}}{24}\mu^s\frac{\Gamma(-s)}{\Gamma(s+N)}
\prod^N_{k=2}\Delta(1-\sqrt{2}p_k)
\left(2\sum^N_{k=2}p_k^2-\sqrt{2}\sum^N_{k=2}p_k-1\right)\,.
\ea
\item[\rm Case 3.]$h=0,\ (-,-,+,\ldots,+)$,
\ba
\lefteqn{A(p_1^{(-)},p_2^{(-)},p_3^{(+)},\ldots,p_N^{(+)})_{h=0}} \nonu
& = & \frac{1}{2}(-1)^{N+1}\mu^{\sqrt{2}(|p_1|+|p_2|)+2-N}
\prod^N_{k=1}\Delta(1-\sqrt{2}|p_k|) \nonu
&   & \times \biggl[\,\prod^{N-3}_{i=1}(\sqrt{2}|p_1|-i)
+\frac{1}{\sqrt{2}}\sum^{N-3}_{k=1}\prod^{k-1}_{i=1}(\sqrt{2}|p_1|-i)
\prod^{N-3-k}_{j=1}(\sqrt{2}|p_2|-j) \nonu
&   & \times \sum_{3\leq n_1<\cdots <n_k\leq N}
\left|
	\sum^k_{i=1}p_{n_i}+p_1
\right|
 +\ (p_1\lra p_2)\ \biggr]\,.
\ea
\end{description}

%%%%%%%%%%%%%%  General factorization of tachyon amplitudes %%%%%%%%%%%

The form of the amplitude (\ref{eqn:dmp-formula}) is suitable for
studying the factorization of tachyon amplitudes.
Let $p_l$ be the momentum parameter of a tachyon operator
with positive chirality.
As was discussed previously, we show the factorization of
the $N$-point tachyon amplitude by taking the limit of $p_l\ra 1/\sqrt{2}$.
To do this end, we use the relation
\be
[\AL{\beta},S\AL{-n}S^{-1}]=
\frac{in}{\mu\beta}S\AL{-n+\beta}S^{-1}\,,
\label{eqn:comm-formula}
\ee
which is derived from the formula
\be
S\AL{-n}S^{-1} = \sum_m R_{p_m}R^*_{n/\sqrt{2}\beta-p_m}
                 \PSI{m}\BPSI{-n}{+m}\,.
\ee
In (\ref{eqn:comm-formula}), we assumed without loss of generality that
$\beta$ is integral-valued.
Applying (\ref{eqn:comm-formula}) repeatedly to (\ref{eqn:dmp-formula}),
we evaluate the residue of the pole at $p_l = 1/\sqrt{2}$
\ba
\lefteqn{A(\ob{p_1}^{(-)},\ldots,\ob{p_m}^{(-)},
  \ob{p_{m+1}}^{(+)},\ldots,\ob{p_N}^{(+)})} \nonu
& \stackrel{p_l\ra 1/\sqrt{2}}{\sim} &
  \lim_{\beta\ra\infty}(-1)^N\pi^{N-3}\frac{1}{\beta}
  \prod^N_{\stackrel{\sst j=1}{j\neq l}}
  \frac{\Gamma(-n_j/\beta)}{\Gamma(n_j/\beta)}
  \mu^{\frac{1}{2\beta}\sum^N_{j=1}n_j}\frac{-1}{1-n_l/\beta} \nonu
&    \!\!& \times \sum^m_{k=1}
  \bra{0}\AL{n_N}\cdots\AL{n_{l+1}}\AL{n_{l-1}}\cdots\AL{n_{m+1}}
  S\AL{-n_m}\cdots\AL{-n_{k+1}}S^{-1} \nonu
&    \!\!& \times [\AL{\beta},S\AL{-n_k}S^{-1}]
  S\AL{-n_{k-1}}\cdots\AL{-n_1}\ket{0} \nonu
& = \!\!& -\frac{\pi}{1-\sqrt{2}p_l}\sum^m_{k=1}
  \frac{\theta(-1-\sqrt{2}p_k)}{1+\sqrt{2}p_k} \nonu
&   \!\!&  \times A(\ob{p_1}^{(-)},\ldots,\!\ob{p_k+1}^{(-)},
  \ldots,\!\ob{p_m}^{(-)},\!
  \ob{p_{m+1}}^{(+)},\ldots,\!\ob{p_{l-1}}^{(+)},\!
  \ob{p_{l+1}}^{(+)},\ldots,\!\ob{p_N}^{(+)}).
\label{eqn:factorization}
\ea
We now have a sum of $m$ tachyon amplitudes, since we observe in
(\ref{eqn:factorization}) the short-distance singularity between
$\ob{p_l}^{(+)}$ and each tachyon operator with negative chirality.
The reason for the appearance of step functions is
that an annihilation mode comes to the right-hand side of the scattering
matrix $S$.
Eq.(\ref{eqn:factorization}) is the result of a sum over all
the topologies.
By expanding both sides of (\ref{eqn:factorization}) with respect to
$\mu^{-1}$, we have a factorization like that in (\ref{eqn:factorization})
for each value of the genus.
This factorization is indeed consistent with the results of Sakai and Tanii.
We can also show a similar factorization by taking a limit of
$p_k\ra -1/\sqrt{2}$ for a tachyon operator ${\cal{T}}_{p_k}$ with
negative chirality.

As a consequence, we obtain the following rules for the factorization of
tachyon amplitudes.
\begin{enumerate}
\item There are no short-distance singularities in an operator
      product expansion between tachyon operators
      with the same chirality.
\item The pole of a physical tachyon state that is produced by
      an operator product expansion of two tachyon operators with
      opposite chiralities is of the first order.
      The chirality of the intermediate tachyon is opposite to that of the
      tachyon operator whose momentum approaches $\pm 1/\sqrt{2}$
      in the limit.
\item The form of the factorization is genus-independent.
\end{enumerate}
We can understand these rules from the point of view of the operator product
expansion of two tachyon operators as in (\ref{eqn:OPE}).
\par
\vspace{3mm}
\par
We would like to thank N. Sakai for a useful discussion on the factorization.
We are grateful to P. Crehan for reading the manuscript.
\vspace{3mm}
%
%%%%%%%%%%%%%%%%%%%%% References %%%%%%%%%%%%%%%%%%%%%%%%%%
%

%

\begin{thebibliography}{99}
%
\bibitem{BIPZ} E. Br\'ezin, C. Itzykson, G. Parisi, and J.B. Zuber,
           {\it Commun.~Math.~Phys.~}{\bf 59} (1978) 59;
           C. Itzykson and J.B. Zuber,
           {\it J.~Math.~Phys.~}{\bf 21} (1980) 411;
           D. Bessis, C. Itzykson, and J.B. Zuber,
           {\it Adv.~Appl.~Math.~}{\bf 1} (1980) 109;
           M.L. Mehta,
           {\it Comm.~Math.~Phys.~}{\bf 79} (1981) 327;
           V.A. Kazakov, I.K. Kostov, and A.A. Migdal,
           {\it Phys.~Lett.~}{\bf 157B} (1985) 295;
           F. David, {\it Nucl.~Phys.~}{\bf B257} (1985) 45;
           V.A. Kazakov, {\it Phys.~Lett.~}{\bf 119A} (1986) 140;
           D.V. Boulatov and V.A. Kazakov,
           {\it Phys.~Lett.~}{\bf 186B} (1987) 379;
           V.A. Kazakov and A.A. Migdal,
           {\it Nucl.~Phys.~}{\bf B311} (1988) 171.
\bibitem{GRMI}
           E. Br\'ezin and V.A. Kazakov,
           {\it Phys.~Lett.~}{\bf 236B} (1990) 144;
           D.J. Gross and A.A. Migdal,
           {\it Phys.~Rev.~Lett.~}{\bf 64} (1990) 127; 717,
           {\it Nucl.\ Phys.\ }{\bf B340} (1990), 333;
           M. Douglas and S. Shenker,
           {\it Nucl.~Phys.~}{\bf B335} (1990) 635.
%
\bibitem{DDK} F. David,
        {\it Mod.\ Phys.\ Lett.\ }{\bf A3} (1988) 1651;
        J. Distler and H. Kawai,
        {\it Nucl.\ Phys.\ }{\bf B321} (1989) 509;
        J. Distler, Z. Hlousek, and H. Kawai,
        {\it Int.\ J. of Mod.\ Phys.\ }{\bf A5} (1990) 391; 1093.
\bibitem{SEIREV} N. Seiberg,
        {\it Prog.\ Theor.\ Phys.\ Suppl.\ }{\bf 102} (1990) 319.
\bibitem{POLREV} J. Polchinski,
        published in Strings '90, Texas A\&M, Coll. Station Workshop (1990) 62;
        {\it Nucl.\  Phys.\ }{\bf B357} (1991) 241.
\bibitem{GOULI} M. Goulian and M. Li,
        {\it Phys.\ Rev.\ Lett.\ }{\bf 66} (1991) 2051;
           A. Gupta, S. Trivedi, and M. Wise,
           {\it Nucl. Phys.\ }{\bf B340} (1990) 475.
\bibitem{DFKU} P. Di Francesco and D. Kutasov,
        {\it Phys.\ Lett.\ }{\bf 261B} (1991) 385;
        {\it Nucl.\ Phys.\ }{\bf B375} (1992) 119;
        Y. Kitazawa, {\it Phys.\ Lett.\ }{\bf 265B} (1991) 262;
        N. Sakai and Y. Tanii,
        {\it Prog.\ Theor.\ Phys.\ }{\bf 86} (1991) 547;
        V.S. Dotsenko, Paris preprint PAR--LPTHE 91--18 (1991).
\bibitem{SATAPART}
           N. Sakai and Y. Tanii,
           {\it Int.\ J. of Mod.\ Phys.\ }{\bf A6} (1991) 2743;
           N. Sakai,
           Tokyo Inst. of Tech. preprint TIT/HEP-205 (1992);
           M. Bershadsky and I.R. Klebanov,
           {\it Phys.\ Rev.\ Lett.\ }{\bf 65} (1990) 3088;
           {\it Nucl.\ Phys.\ }{\bf B360} (1991) 559.
\bibitem{SATAFACT} N. Sakai and Y. Tanii,
        {\it Phys.\ Lett.\ }{\bf 276B} (1992) 41;
        {\it Prog.\ Theor.\ Phys.\ Supp.\ }{\bf 110} (1992) 117.
\bibitem{DMP} R. Dijkgraaf, G. Moore, and R. Plesser,
        Princeton preprint IASSNS--HEP--92/48 (1992).
\bibitem{LIAN} B.H. Lian and G.J. Zuckerman,
          {\it Phys.\ Lett.\ }{\bf 254B} (1991) 417;
          {\bf 266B} (1991) 21;
          S. Mukherji, S. Mukhi, and A. Sen,
          {\it Phys.\ Lett.\ }{\bf 266B} (1991) 337;
          P. Bouwknegt, J. McCarthy, and K. Pilch, CERN preprint
          CERN--TH--6162--91 (1991);
          N. Ohta, Osaka preprint OS-GE 26-92 (1992).
\bibitem{MOORE} G. Moore,
         {\it Nucl.\ Phys.\ }{\bf B368} (1992) 557;
         G. Moore, R. Plesser, and S. Ramgoolam,
         {\it Nucl.\ Phys.\ }{\bf B377} (1992) 143.
%
\end{thebibliography}
\end{document}